\documentstyle[aps]{revtex}
\begin{document}

\twocolumn[\hsize\textwidth\columnwidth\hsize\csname %
@twocolumnfalse\endcsname

\rightline{Submitted to the Physical Review B}
\bigskip

\title{
Scaling Regimes, Crossovers, and Lattice Corrections
in 2D Heisenberg Antiferromagnets}

\author{Norbert Elstner$^*$, Rodney L. Glenister and Rajiv R.P. Singh}
\address{Department of Physics, University of California,
Davis, CA 95616}
\author{Alexander Sokol}
\address{Department of Physics and Frederick Seitz
Materials Research Laboratory, \\
University of Illinois at Urbana-Champaign,
Urbana, IL 61801\\
and L.D. Landau Institute, Moscow, Russia}

\date{\today}

\maketitle

\phantom{.}

\phantom{.}

We study scaling behavior in 2D,
square lattice, $S=1/2$ and $S=1$ Heisenberg antiferromagnets
using the data on full q-dependences of the
equal time structure factor and the static
susceptibility, calculated
through high temperature series expansions.
We also carry out comparisons with a model of two coupled S=1/2
Heisenberg planes with the interlayer exchange
coupling tuned to the $T=0$
critical point (two-plane model hereafter). For both S=1/2 and S=1 models,
we separately determine the spin-wave velocity $c$
and mass $m=c/\xi$, in addition to the correlation length, $\xi$, and
find that $c$ is temperature dependent;
only for temperatures below
$T\alt JS$, where $J$ is the exchange coupling, $c$ approaches its
known $T=0$ value, $c_0$.
This non-universal lattice effect is
caused by the quantum nature of spin, and is therefore
not captured by the quantum nonlinear $\sigma$-model.
Despite this temperature dependence of the spin-wave velocity, full
$q-$ and $\omega$-dependences of the dynamical susceptibility
$\chi(\bf q,\omega)$ agree with those of the universal scaling function,
computable for the $\sigma$-model,
for temperatures up to $T_0 \sim 0.6 c_0/a$, and
their further analysis leads
us to the inference that below $T_0$
the S=1 model is
in the renormalized classical (RC) regime, the two-plane model
is in the quantum critical (QC) regime in agreement with earlier work,
and the S=1/2 model exhibits a RC-QC crossover,
centered around T=0.55J. In particular, for the
S=1/2 model above the RC-QC crossover
and for the two-plane model at all temperatures where calculated,
the obtained spin wave mass $m=c/\xi$ is in excellent agreement with
the known universal QC prediction,  $m\simeq 1.04\, T$.
In contrast, for the S=1/2 model below the RC-QC crossover, and for
the S=1 model at all temperatures, the behavior agrees with the exact
RC expression.
For all three models,
nonuniversal behavior occurs above $T\sim 0.6c_0/a$.
Our results
strongly support the conjecture by Chubukov and
Sachdev that the $S=1/2$ model is close enough to the $T=0$
critical point to exhibit QC behavior at intermediate temperatures.

\phantom{.}

\phantom{.}

$^*$Present address: Service de Physique Th\'eorique,
CEA-Saclay,
91191 Gif-sur-Yvette Cedex,
France.
]

\pagebreak

\phantom{.}

\newpage

\phantom{.}

\newpage

\tableofcontents

\section{Introduction} \label{introduction}

The problem of antiferromagnetism in two dimensions has recently
attracted much attention, in part due to its relevance
to high temperature superconductivity. Since the high
temperature
superconducting materials are derivatives, obtained by doping,
of parent insulating layered S=1/2
antiferromagnets, theory of the insulating
phase is important for understanding the unusual magnetic and
transport properties in the normal state of doped materials.
Our results are directly relevant to the following two materials:
$\rm La_2CuO_4$, which is the insulating
parent compound of the  $\rm La_{2-x}Sr_xCuO_4$ and
$\rm La_{2-x}Ba_xCuO_4$ superconductor families,
and $\rm Sr_2CuO_2Cl_2$,
which is unrelated to any
superconducting family, but has well studied
magnetic properties, similar
to those of $\rm La_2CuO_4$. Both materials experience
antiferromagnetic transitions to the Neel state with nonzero staggered
magnetization, $\rm La_2CuO_4$ at $T_N\simeq 300K$ and
$\rm Sr_2CuO_2Cl_2$ at $T_N\simeq 256K$.
According to the Hohenberg-Mermin-Wagner theorem,
in genuine two dimensional
antiferromagnets the long range order at finite temperatures
is prevented by thermal fluctuations. In real materials,
small interlayer exchange interaction allows the
transition to occur at a finite temperature $T_N$.
In what follows we discuss only the properties well above $T_N$,
where the magnetic subsystem can be considered two-dimensional.

It is well established \cite{reviews} that the magnetic properties of
$\rm La_2CuO_4$ and $\rm Sr_2CuO_2Cl_2$
are described by the $S=1/2$ Heisenberg model
on the two-dimensional square lattice. The Heisenberg model
is defined by the following Hamiltonian:
\begin{equation}
H = J \sum_{\langle ij\rangle} {\rm \bf S}_i {\rm \bf S}_j,
\end{equation}
where ${\rm \bf S}_i$ are local spin operators and the notation
$\sum_{\langle ij\rangle}$ means that the sum is taken over
all pairs of nearest neighbors.
In what follows we set $k_B = \hbar = a = 1$, where $a$ is the
lattice constant, although in certain expressions we retain $a$ for
clarity.
The dynamical spin susceptibility
$\chi({\bf q},\omega)=\chi'+i\chi''$, is
defined such that the magnetization induced by a wavevector and frequency
dependent magnetic field is:
\begin{equation}
{\bf M}({\bf q},\omega) =
\chi({\bf q},\omega) {\bf H}_{\rm ext}({\bf q},\omega).
\end{equation}
We calculate high temperature
series expansions in powers of $\beta=J/T$ for two quantities:
the static q-dependent susceptibility, defined as
a response to a static, but generally nonuniform, magnetic field,
\begin{equation}
\chi({\bf q}) \equiv \chi({\bf q},\omega=0),
\end{equation}
and the Fourier transform of the
equal time correlation function of two spin operators,
\begin{equation}
S({\bf q}) = \sum_r \langle {\rm S}_z({\bf 0})
{\rm S}_z({\bf r}) \rangle \exp(i{\bf qr}),
\end{equation}
where angular brackets represent thermal averages.
These quantities can be expressed in terms of $\chi''({\bf q},\omega)$,
which plays the role of the spectral weight for spin fluctuations,
according to:
\begin{eqnarray}
S({\bf q}) = T \sum_{n=-\infty}^{+\infty} \chi({\bf q},i\omega_n) =
\frac{1}{2\pi} \int_{-\infty}^\infty d\omega \,
\frac{\chi''({\bf q},\omega)}{\mbox{th}(\omega/2T)},
\nonumber
\\
\chi({\bf q}) =  \chi({\bf q},i\omega_n=0)  =
\frac{1}{\pi} \int_{-\infty}^\infty d\omega \,
\frac{\chi''({\bf q},\omega)}{\omega}
\label{spectral}
\end{eqnarray}
where $\omega_n = 2\pi nT$
is the imaginary bosonic Matsubara frequency.

Total spin conservation in the model leads to the relation
\begin{equation}
\chi_0 = g^2\mu_B^2 \chi({\bf q}=0) = g^2\mu_B^2 T^{-1} S({\bf q}=0),
\end{equation}
where $\chi_0$ is the bulk magnetic susceptibility, $g$ the
electron gyromagnetic ratio, and $\mu_B$ the Bohr magneton.

We study both the case of $S=1/2$, which corresponds to
$\rm La_2CuO_4$ and $\rm Sr_2CuO_2Cl_2$, and the case of $S=1$.
We also study a model of two S=1/2 coupled Heisenberg planes
at its critical point, $J_2/J_1\simeq 2.51\pm 0.02$
\cite{Sandvik:Scalapino:PRL}.
For both $\chi({\bf q})$ and $S({\bf q})$,
we are able to generate
series for an arbitrary wavevector ${\bf q}$
complete to order $\beta^{14}$
for the $S=1/2$ model, and to $\beta^{10}$ for the $S=1$ model.
For the two-plane model we set $J_2/J_1=2.5$, and generate
series for the structure factor and the susceptibility at the
antiferromagnetic ordering vector to order $\beta^{10}$.
We use Pade approximation techniques for series analysis.
We emphasize that the series converges extremely well in the
temperature region presented in the paper.
In fact, with the available long series,
most of the Pade approximants
are within a per cent, or so, from each other,
down to $T\approx 0.4J$ for $S=1/2$ and $T\approx 1.5J$
for $S=1$.
This allows us to put quite small errorbars on our results for
all but the lowest temperatures. At the lowest temperatures, the
approximants begin to significantly differ from each other and hence we
do not go to still lower temperatures.

\section{Overview of Renormalized Classical
versus Quantum Critical Scaling Behavior} \label{RC.vs.QC}

Scaling behavior of the 2D Heisenberg models follows from their mapping
\cite{CHN} to the quantum nonlinear sigma model ($\sigma$-model hereafter).
When the long range order is present at $T=0$,
fluctuations at low enough temperatures
are purely classical \cite{CHN}, in other words the
characteristic frequency of spin fluctuations $\bar{\omega}\ll T$.
This happens because
for $T\ll \rho_s$ ($\rho_s$ is the T=0 spin stiffness),
\begin{equation}
\bar{\omega} \sim T^{1/2} \exp(-2\pi \rho_s/T)
\end{equation}
decreases faster than the temperature as the temperature
decreases.
Accordingly, this regime is called the
{\em renormalized classical} (RC)
regime, where ``renormalized'' means that the $T=0$ parameters, such as
the spin stiffness, $\rho_s$, and spin wave velocity, $c$, are renormalized
compared to their mean-field values as a result of quantum fluctuations
at shorter wavelengths.

Recently, an asymptotic expression for the correlation length
has been derived for this case including the exact value of the prefactor
\cite{Hasenfratz:Niedermayer}:
\begin{equation}
\xi = \frac{e}{8} \frac{c}{2\pi \rho_s}
\exp\left(\frac{2\pi\rho_s}{T}\right)
\left( 1 - \frac{T}{4\pi \rho_s} +
O\left[\frac{T^2}{\rho_s^2}\right] \right),
\label{HN}
\end{equation}
where $e$ is the natural logarithm base.
Here $\xi$ is defined such that the dominant behavior
of the  pair correlation function
at large distances is $\exp(-r/\xi)$ with  power-law corrections.

The {\em quantum critical}
(QC) regime is the high temperature scaling regime in the
phase diagram of the $\sigma$-model, derived by Chakravarty, Halperin and
Nelson \cite{CHN},
where the dominant energy scale is set by the temperature:
\begin{equation}
\bar{\omega} \sim T
\end{equation}
The crossover between RC and QC regimes occurs at
$T\sim \mbox{const} \times \rho_s$.

As the temperature increases even further, the assumptions behind
the mapping of the Heisenberg model to the $\sigma$-model eventually
become invalid, which at temperatures $T\agt c/a$
would destroy scaling altogether.

Chubukov and Sachdev \cite{CSY} recently argued
that crossover from RC to QC regime
occurs around $T=0.4J$ for the S=1/2 model.
According to their paper,
in the region of $T=0.4J-0.6J$, the system
exhibits QC behavior.
The conclusions in \cite{CSY} were based on the agreement
of quantum critical scaling predictions for
the bulk susceptibility, $\chi_0$,
and the NMR spin-lattice relaxation rate, $1/T_1$, with the
experimental measurements \cite{Johnston,Imai} in $\rm La_2CuO_4$ and with
numerical calculations for the Heisenberg model.

However, the temperature dependence of the correlation length did not
appear to support this picture of the QC behavior at intermediate
temperatures \cite{Greven}.
$\xi(T)$ has been measured by means of
inelastic neutron scattering in both
$\rm La_2CuO_4$ \cite{Keimer} and $\rm Sr_2CuO_2Cl_2$ \cite{Greven},
and calculated
numerically for the 2D Heisenberg model using Monte-Carlo simulations
\cite{Ding:Makivic} and series expansions approaches
\cite{Singh:Glenister}. All experimental measurements
and numerical results agree well with each other, and the RC expression
seems to describe well the combined experimental and numerical
data for the correlation length not only for $T\alt\rho_s$ \cite{CHN},
but even
for temperatures as high as $T=J\sim 5\rho_s$ \cite{Greven}.
At $T=J$, the second term in the expansion in powers of
$T/\rho_s$ is only about two times smaller
than the first term, therefore
Eq.(\ref{HN}) may remain accurate to all orders only if the
coefficients in front of higher terms in the series in $T/\rho_s$
are anomalously small. Since these coefficients are
presently unknown and may indeed turn out to be small, on the basis of
the correlation length data alone one would be tempted to conclude that
the system remains in the RC regime for
temperatures up to $T=J$, which would also imply an absence of
a QC regime for this model \cite{Greven}.

The main thesis of the present paper is that such a
conclusion is not justified if one examines the detailed
wave-vector and temperature dependences of
$S({\bf q})$ and $\chi({\bf q})$.
It has been shown
in our earlier paper \cite{Sokol:Glenister:Singh} that the scaling
functions for the static susceptibility $\chi({\bf q})$, and equal
time correlation function, $S({\bf q})$, in the temperature range
$T=0.6J-J$ are that of the $\sigma$-model in the QC rather
than the RC regime. In that temperature range, we also obtained
various universal dimensionless ratios in excellent agreement with their
respective QC predictions,
with the RC predictions being far outside
the errorbars.

However, the temperature dependence of the
correlation length still presented a puzzle.
In the quantum critical region, the leading temperature dependence
of the correlation length is of
the form
\begin{equation}
\xi=X^{-1}_\infty \frac{c}{T},
\label{xi:QC}
\end{equation}
where $c$ is the spin-wave velocity and $X_\infty$ is a universal constant.
With the known value of $X_\infty$
\cite{CHN,CSY,Manousakis:Salvador}
and the known zero temperature
value of the spin-wave velocity,
the temperature dependence of the
correlation length did not fit Eq.(\ref{xi:QC}) in any
extended temperature region, although
the magnitude of $\xi$ around $T\sim J/2$ is in agreement with
Eq.(\ref{xi:QC}).
In Ref. \cite{Sokol:Glenister:Singh},
we conjectured that there are
nonuniversal lattice corrections at intermediate temperatures,
which affect such quantities as the correlation length, but nevertheless
the scaling functions and the characteristic frequency of spin fluctuations
are that of the QC regime.

In the present paper, we are able to
show that the disagreement of the correlation length
with the quantum critical prediction is indeed a result of lattice
corrections, more specifically corrections to the spin wave velocity, which
is temperature dependent especially at higher temperatures.
We further show that when Eq.(\ref{xi:QC})
is evaluated using the correct $c(T)$ rather than its $T=0$ value,
the agreement with the QC behavior is observed for temperatures above
$T\sim 0.55J$.

\section{Universal Scaling Functions} \label{universal}

The first stage of our analysis is to establish whether or not,
and in what temperature range, universal
$q-$ and $\omega-$dependences of $\chi({\bf q},\omega)$, derived for the
quantum nonlinear $\sigma$-model, apply to the S=1/2 and S=1 square lattice
Heisenberg antiferromagnets. Here,
it is convenient to introduce the spin wave mass defined as
\begin{equation}
m(T) = \frac{c(T)}{\xi(T)},
\end{equation}
which, in what follows, replaces $\xi(T)$ as a scaling parameter.
We emphasize that both $c(T)$ and $\xi(T)$ are measurable
thermodynamic quantities.
This replacement facilitates scaling comparisons:
we will show that when at temperatures $T\sim JS$
the spin wave velocity becomes temperature dependent
due to lattice corrections to $c$,
the behavior of $m$ remains universal.

For the $\sigma$-model calculations,
we employ $1/N$ expansion approach and utilize
some of the results obtained earlier by Chubukov and Sachdev
\cite{CSY}. We do not calculate the frequency dependence of the
dynamical susceptibility explicitly.
Instead, we base our comparisons
on the calculated difference between the equal time correlators
and static susceptibilities
at all wavevectors, which is a sensitive probe of
the frequency distribution of the spectral weight.
Our calculations are discussed in detail in Appendix
\ref{sigma:app}, where we derive the following equations which relate
$S({\bf q})$ and $\chi({\bf q})$, computable
via series expansions for the Heisenberg
models, to the scaling parameters $m$ and $c$:
\begin{equation}
\frac{\chi({\bf Q})}{\chi({\bf q})} =
1 + \left( \frac{c}{m} \right)^2 |{\bf q}-{\bf Q}|^2,
\label{chi:lin:3}
\end{equation}
\begin{equation}
\gamma^2 \left[ \frac{S({\bf q})}{T\chi({\bf q})} \right] =
\left( u_2 \frac{m}{T} \right)^2 + \left( \frac{c}{T} \right)^2
|{\bf q}-{\bf Q}|^2
\label{Schi:lin:3}
\end{equation}
where
\begin{equation}
u_2 \simeq 1 + 0.1473/N = 1.0491
\label{uv}
\end{equation}
and the function $\gamma[x]$ is defined according to:
\begin{equation}
\gamma\left[\frac{x/2}{\tanh(x/2)}\right] \equiv x,
\ \ \ \ {\rm for} \ \ \ \ x>1.
\label{gamma:def:1}
\end{equation}
Here the expression for $u_2$ is evaluated at $N=3$,
which corresponds to the physical three-component spin.
As discussed in Appendix \ref{sigma:app},
these equations are not exact. In particular, there
are nonzero higher order terms in $|{\bf q}-{\bf Q}|^2$
and multiplicative factors in front of $c$.
However, such corrections are found to be
less than $1-2\%$, which greatly simplifies further considerations.

Eqs.(\ref{chi:lin:3},\ref{Schi:lin:3}) are in
a form that greatly facilitates comparison with the numerical data: the
left hand parts of the above equations are computable numerically for the
Heisenberg models, and their linear dependence on
$|{\bf q}-{\bf Q}|^2$, expected from Eqs.(\ref{chi:lin:3},\ref{Schi:lin:3}),
can be verified by plotting on the
appropriate scale.
In the temperature range where the expected linear
behavior is indeed observed, the scaling parameters $m$ and $c$ can
be determined independently from the intercept and slope of the plotted data.

Before we proceed with the comparisons, an important remark is in order.
Since at high temperatures
$\chi({\bf q})$ and $S({\bf q})$ do not exhibit strong $q$-dependence,
their small deviations from the values at ${\bf Q}=(\pi,\pi)$
are always linear in $|{\bf q}-{\bf Q}|^2$,
as required by the lattice symmetry, so that at
high enough temperatures $q$-dependences in
Eqs.(\ref{chi:lin:3},\ref{Schi:lin:3}) would convey no information about
scaling. The high temperature crossover to non-universal weakly
interacting spins is discussed more thoroughly in the next section.

\begin{figure}
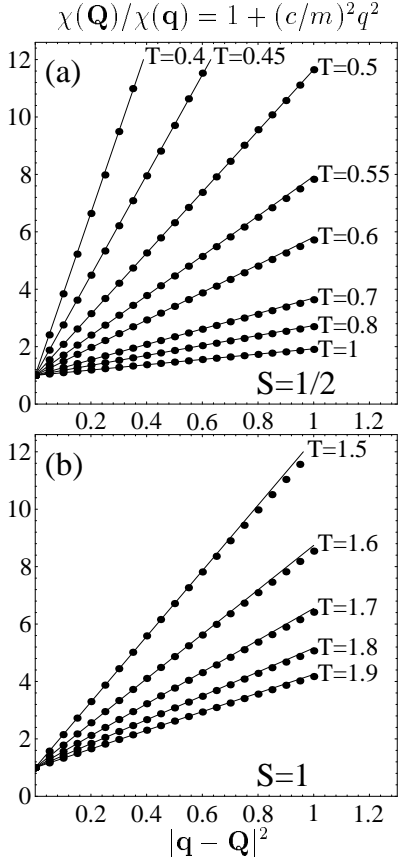

\caption{
$\chi({\bf Q})/\chi({\bf q})$ is plotted as a function
of $|{\bf q}-{\bf Q}|^2$ for the S=1/2 (a) and S=1 (b) Heisenberg
models for selected temperatures.
Linear dependences, in agreement
with those derived for the $\sigma$-model
Eq.(\protect\ref{chi:lin:3}), are observed.}
\label{chi:fig}
\end{figure}

We start with Eq.(\ref{chi:lin:3})
and plot its left hand side, calculated
using high temperature series expansions,
versus $|{\bf q}-{\bf Q}|^2$
for the S=1/2 and S=1 Heisenberg models in Fig.\ref{chi:fig}.
Eq.(\ref{chi:lin:3}) predicts that the plotted data should be linear.
For wavevectors
$|{\bf q}-{\bf Q}| \le 0.8$, this is indeed the case for both S=1/2
and S=1 models in a broad range of temperatures: $T=0.4J-J$ for
S=1/2, and $T=1.5J-1.9J$ for S=1.

\begin{figure}
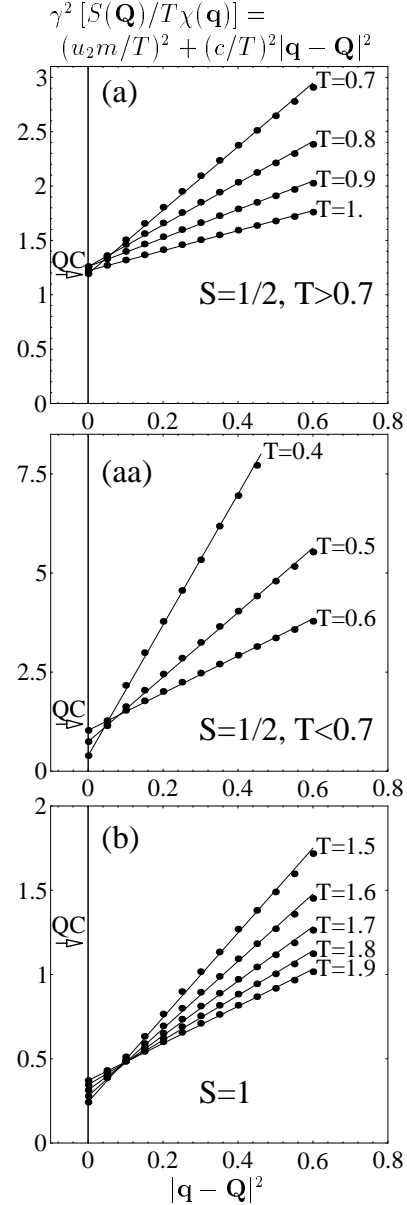

\caption{
$\gamma^2\left[S({\bf Q})/T\chi({\bf q})\right]$
is plotted as a function
of $|{\bf q}-{\bf Q}|^2$ for the S=1/2 (a,aa) and S=1 (b) Heisenberg
models for selected temperatures.
The dependence is linear, in agreement
with those derived for the $\sigma$-model
Eq.(\protect\ref{Schi:lin:3}). For the S=1/2 model
the intercept at
$|{\bf q}-{\bf Q}|=0$ is
nearly temperature independent above $T=0.6J$, and is in agreement
with the known universal QC prediction $\simeq 1.19$, shown as
an arrow. For temperatures lower than $T=0.5J$
for S=1/2, and for all temperatures
for S=1, the intercept is smaller
than the QC prediction and decreases as the temperature decreases,
indicating crossover to RC behavior.}
\label{Schi:fig}
\end{figure}

We now turn to Eq.(\ref{Schi:lin:3}) and plot its
left hand side versus $|{\bf q}-{\bf Q}|^2$ in Fig.\ref{Schi:fig}.
Again, as
expected from Eq.(\ref{Schi:lin:3}), the plotted data
is linear in the same wavevector and temperature range for both S=1/2
and S=1 models.

Both Figs.\ref{chi:fig},\ref{Schi:fig} are generated with ${\bf q}-{\bf Q}$
pointing along the diagonal of the Brillouin zone. The data with ${\bf
q}-{\bf Q}$ pointing along other directions is virtually identical
in the same wavevector and temperature range, as expected in the
isotropic $\sigma$-model. Further,
a straightforward estimate shows that corrections to linear spin wave
spectrum due to higher gradients, not included in the $\sigma$-model, are
only about 5\% for the wavevector range of interest,
$|{\bf q}-{\bf Q}|<0.8$.
The studied temperature range
is limited from below by our ability to
evaluate accurately $\chi({\bf q})$ and $S({\bf q})$ from our
high temperature series, presently known to $\beta^{14}$ for the S=1/2
model, and to $\beta^{10}$ for the S=1 model.

{}From the slope and intercept of the linear fits to the data plotted
in Fig.\ref{Schi:fig}, we are able to determine
$c$ and $m$ separately. Since this data is used extensively
in what follows, we carry out comparisons of our results for $m$ 
and $c$ with direct calculations of the correlation length.
The comparisons, which are described in Appendix \ref{xi:app},
confirm that our calculations are indeed accurate.

\begin{figure}
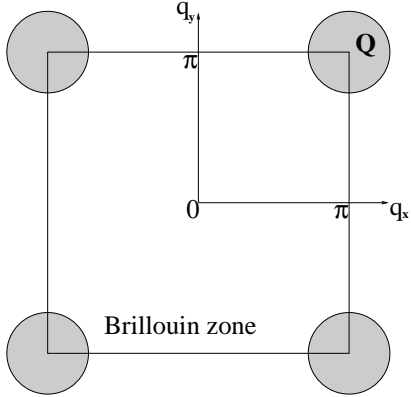

\caption{
A drawing of the Brillouin zone for the square lattice; shaded area shows
the wavevector range where spin correlations of the Heisenberg models are
described by the $\sigma$-model.}
\label{brillouin:fig}
\end{figure}

We have thereby established that Eqs.(\ref{chi:lin:3},\ref{Schi:lin:3})
are valid in the range $T=0.4J-J$ ($T=1.5J-1.9J$)
for S=1/2 (S=1) for $|{\bf q}-{\bf Q}| \le 0.8$, which means that
wavevector dependences of $S({\bf q})$ and $\chi({\bf q})$ in the
Heisenberg models are described by the $\sigma$-model in the same
wavevector and temperature range, which is shaded in
Fig.\ref{brillouin:fig}.

\section{Scaling Regimes} \label{regimes}

\subsection{Quantum Nonlinear $\sigma$-Model} \label{QNLSM:scaling}

We now proceed with the discussion of scaling regimes.
The dynamical scaling form of the
staggered correlator is \cite{CHN}:
\begin{equation}
\chi_s(q,\omega) = \chi_s(0) \,
 \Phi^{(i)}\left(\frac{cq}{m},\frac{\omega}{\bar{\omega}}\right),
\label{scaling:chi}
\end{equation}
where the scaling function for the renormalized classical
regime, $\Phi^{\rm RC}$, differs from that
in the quantum critical
regime, $\Phi^{\rm QC}$. In general, $\bar{\omega}$ and
$m$ are not equivalent:
the spin wave velocity, and therefore $m$, are defined at $q\gg
\xi^{-1}$, while $\bar{\omega}$ is defined at $q\sim \xi^{-1}$.
In the quantum 
nonlinear $\sigma$-model, $c$ is expected to be
temperature dependent only when the temperature is comparable
to the ultraviolet cutoff  $\Lambda$
(except in case of relativistic invariance of the cutoff
procedure, when
$c$ is always temperature independent and coincides with its bare
value).

In the RC regime temperature dependences of the spin wave mass
for $N=\infty$ and $N=3$ are different:
\cite{CHN,CSY}:
\begin{equation}
m \sim  \exp \left( -\frac{2\pi\rho_s}{T} \right) \times
\left\{
\begin{array}{cl}
T      &\mbox{ for $N=\infty$} \\
\rho_s &\mbox{ for $N=3$}
\end{array}
\right. .
\end{equation}
Furthermore, in this regime
$\bar{\omega}$ is not equivalent to $m$, except for $N=\infty$:
\begin{equation}
\frac{\bar{\omega}}{m} \sim
\left\{
\begin{array}{ll}
1               &\mbox{ for $N=\infty$} \\
(T/\rho_s)^{1/2} &\mbox{ for $N=3$}
\end{array}
\right. .
\label{omegabar:RC}
\end{equation}
The exact asymptotic expression for $m$
follows immediately from
Eq.(\ref{HN}):
\begin{equation}
m = \frac{16\pi \rho_s}{e}
\exp\left(\frac{-2\pi\rho_s}{T}\right)
\left( 1 - \frac{T}{4\pi \rho_s} +
O\left[\frac{T^2}{\rho_s^2}\right] \right)^{-1}.
\label{m:RC}
\end{equation}
The value of $\rho_s$, which is the only dimensional parameter in
Eq.(\ref{m:RC}), is known for the Heisenberg models from
zero temperature series expansions \cite{Singh:rhos}, quantum Monte
Carlo simulations \cite{Wiese:Ying}, and
$1/S$ calculations \cite{1/S}.

In the QC regime, temperature is the only energy scale in the problem,
so that:
\begin{equation}
\bar{\omega} \sim m \sim T,
\end{equation}
which is valid for both $N=3$ and $N=\infty$ models. Furthermore,
the ratio
$m/T$ is universal in the $\sigma$-model, and has been calculated
via the $1/N$ approach \cite{CSY}: 
\begin{equation}
\frac{m}{T} = \Theta\, \big(1+{0.2373\over N}\big) \sim 0.9624 +
\frac{0.2284}{N} \approx 1.04,
\label{m:QC}
\end{equation}
where $ \Theta = 2 \log[(1+\sqrt{5})/2] $. Given that already the first
subleading $1/N$ correction is small ($\sim 8\%$)
for $N=3$, Eq.(\ref{m:QC}) is likely to be close to the
true value for the $O(3)$ $\sigma$-model, and is hereafter
adopted as the QC prediction. It can be compared with calculations
based on other techniques:
\begin{equation}
\frac{m}{T} = \left\{
\begin{array}{ll}
0.9, & \mbox{ $2+\epsilon$ expansions \cite{CHN}} \\
1.25\pm .25, & \mbox{ Monte Carlo \cite{Manousakis:Salvador} }
\end{array}
\right. .
\label{m:all.methods:QC}
\end{equation}
The results are generally consistent with each other. Further support
for the numerical value obtained in the $1/N$ calculation
comes from the study of the two-plane model at its
$T=0$ critical coupling by
Sandvik and Scalapino \cite{Sandvik:Scalapino:PRL},
where temperature dependence of the correlation
length was found to be in agreement with that derived from
Eq.(\ref{m:QC}).

Thus, both RC (\ref{m:RC}) and QC (\ref{m:QC})
predictions for $m(T)$ are
available to carry out comparisons with
Heisenberg models without any adjustable parameters.

\subsection{Heisenberg Models} \label{heisenberg:scaling}

We start by comparing behavior of the spin wave mass for
S=1/2, S=1, and the two plane models.
The S=1 model, due to its larger spin, is
``more classical'' than the S=1/2 model, hence it is less likely to have
a range of quantum critical behavior. On the other hand, for
the two plane model at its critical point, temperature dependences of
$\chi_0$ and $\xi$ consistent with the QC
regime had been reported earlier by
Sandvik and Scalapino \cite{Sandvik:Scalapino:PRL}.
The comparison of different models proves quite informative.

\begin{figure}
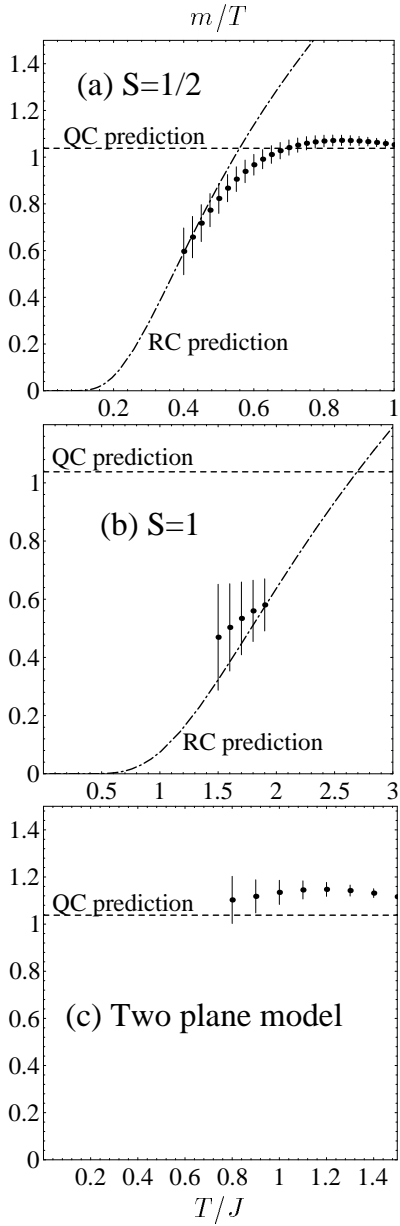

\caption{
The ratio $m/T$ is plotted for (a) the S=1/2, (b) S=1
and (c) the two-plane Heisenberg models
versus $T/J$ along with the known RC
(Eq.(\protect\ref{m:RC}),
dot-dashed line) and QC (Eq.(\protect\ref{m:QC}), dashed line)
universal predictions.
The comparisons do not contain any adjustable parameters,
because the spin stiffness entering the RC expression is known from
earlier studies \protect\cite{Singh:rhos,1/S}.  For
the ``more classical'' S=1 model renormalized classical
behavior, and for the ``critical'' two
plane model quantum critical
behavior, are observed at all temperatures. In contrast,
for the S=1/2 model a
crossover from RC to QC behavior occurs around $T\simeq 0.55J$.}
\label{m:fig}
\end{figure}

We now plot $m(T)$ for these three models
along with the universal RC (\ref{m:RC}) and QC (\ref{m:QC})
scaling predictions in Fig.\ref{m:fig}.
It is evident from Fig.\ref{m:fig} that $m(T)$ for the S=1 model shows RC
behavior at all temperatures, for the two plane model
QC behavior at all temperatures, and for the S=1/2 single plane model a
crossover between RC behavior
below $T\sim 0.45J$ and QC behavior above $T=0.65J$.

\begin{figure}
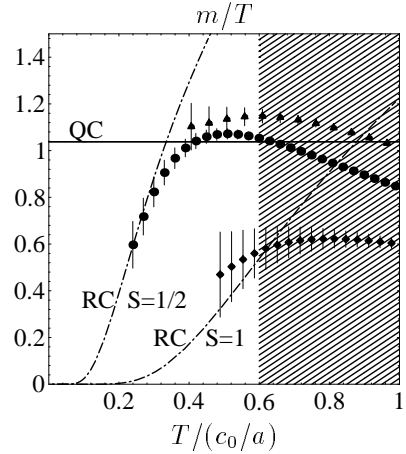

\caption{
$m/T$ is plotted versus the ratio $T/(c_0/a)$ for the
S=1/2 (circles) and S=1 (diamonds) single-plane models,
as well as the model of two S=1/2
planes (triangles)
coupled such that the system is at the critical point
(as determined by Sandvik and Scalapino)
and therefore should exhibit QC behavior at all temperatures.
The difference between models in the temperature range where scaling applies
is evident: the S=1 model is always renormalized classical, the two plane
model is always quantum critical, and the S=1/2 model exhibits a
crossover between the two around $T=0.5J$. Moreover, for S=1/2 and S=1
models $m/T$ deviates from the respective scaling temperature dependences at
roughly the same value of $T/(c_0/a)>0.6$ (shown as shaded area),
as expected given the similar character of deviations from linearity
in their spin wave spectrum (see text). For $c_0$, we use the spin wave
theory results: $c_0=1.67J$ for S=1/2, $c_0=3.07J$ for S=1
\protect\cite{1/S}, and
$c_0=1.9J$ for the two-plane model \protect\cite{Chubukov:Sandvik}.}
\label{combined:fig}
\end{figure}

We now discuss the onset of nonuniversal behavior at high
temperatures. The temperature scale at which
nonuniversal behavior arises is set roughly by $c/a$.
The spin wave spectrum becomes
nonlinear at $q\sim a^{-1}$, and the excitations in this wavevector range
are suppressed only
when $T\ll c/a$. The spin stiffness, on the other hand, is less
relevant to this crossover to nonuniversal behavior at high temperatures.
To a crude approximation we expect scaling to fail for S=1/2 and S=1
models
at $T/(c/a)$ of the same order, given their similar spin wave spectrum.
In order to study failure of scaling, we therefore
plot the ratio $m/T$ versus $T/(c_0/a)$ (Fig.\ref{combined:fig}).
The use of $c_0$ instead of the
temperature dependent $c(T)$ is to avoid a nonlinear temperature
scale. In this particular case, the difference between $c$ and $c_0$
is less important given the ambiguous character of the upper boundary of
scaling behavior.

One can see from Fig.\ref{combined:fig}
that deviations from the universal behavior of $m$
occurs at similar values of the ratio
$T/(c_0/a) \sim 0.6 $ for both S=1/2 and S=1 models,
above which a crossover to the $T\gg J$
asymptotics $m\sim 1/T$ occurs. The region where $m$ becomes nonuniversal
and no longer represents the spin wave mass is shaded 
in Fig.\ref{combined:fig}. Expressed in terms of $J$, the
range where universal behavior of $m(T)$ holds is below
$T\sim J$ for the S=1/2  model and below $T\sim 1.9J$ for the S=1 model.

\begin{figure}
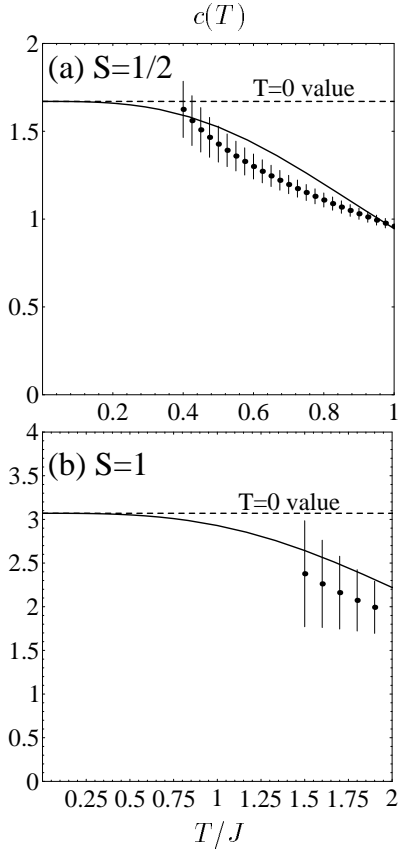

\caption{
The spin wave velocity
plotted as a function of temperature for the S=1/2 (a) and S=1 (b)
Heisenberg models.
Dashed line shows the known T=0 values, $c_0$ \protect\cite{Singh:c,1/S},
and solid line is the result of $1/S$ calculation by Kaganov and Chubukov
\protect\cite{Kaganov:Chubukov}, Eq.\protect\ref{c:kaganov.chubukov}.
The temperature dependence for $T>JS$
arises from nonuniversal lattice corrections.}
\label{c:fig}
\end{figure}

Now, we turn to the analysis of the spin wave velocity calculated in
Sect.\ref{universal}, and plot $c$
determined from the data at different temperatures using
Eq.\ref{Schi:lin:3}
in Fig.\ref{c:fig}.

Dashed lines show the known $T=0$ value of the spin wave velocity calculated
using $1/S$ expansions \cite{1/S},
\begin{eqnarray}
c &= 2\sqrt{2} JS \, (1 + 0.1580/2S + 0.0216/(2S)^2)
\nonumber
\\
 &\simeq \left\{ \begin{array}{ll}
1.67J & \mbox{ for S=1/2} \\
3.07J & \mbox{ for S=1}
\end{array} \right.,
\label{c:1/S}
\end{eqnarray}
which for S=1/2 compares favorably with the T=0 series
expansion calculations \cite{Singh:rhos}.

For all temperatures studied the spin wave velocity turns out
to be {\em temperature dependent}
for both S=1/2 and S=1 models. Towards the
lowest temperatures in the numerically accessible range, $c$
approaches the expected T=0 values given by Eq.(\ref{c:1/S}). Since in the
$\sigma$-model the spin wave velocity is temperature independent except near
the ultraviolet cutoff, which in our case is set by the scale of order
$c/a \sim 2\sqrt{2}JS$, the observed 
temperature dependent spin wave velocity clearly results from the
quantum nature of spin, when the temperature becomes comparable to $JS$.

The quantum corrections to the spin wave velocity
originating from higher order $1/S$ terms in the spin wave theory,
were calculated by Kaganov and Chubukov \cite{Kaganov:Chubukov}:
\begin{equation}
c(T) = c_0 \left[ 1 + \frac{\zeta(3)}{8\pi S}
\left(\frac{T}{SJ}\right)^3 \right]^{-1},
\label{c:kaganov.chubukov}
\end{equation}
where $\zeta(x)$ is the Riemann zeta function. The predictions of
Eq.(\ref{c:kaganov.chubukov}), evaluated for S=1/2 and S=1, are shown as
solid lines in Fig.\ref{c:fig}a,b, respectively. The
agreement with numerical calculations is quite good. Despite
higher order terms in the expansion in powers of $T/JS$, it appears
that Eq.(\ref{c:kaganov.chubukov}) describes the observed temperature
dependence of $c$ fairly well for both $S=1/2$ and $S=1$.

At this stage an important question is,
does the temperature dependence of $c$ destroy scaling altogether?
One might expect that once lattice corrections become important
for quantities like the spin-wave velocity,
they might also lead to a breakdown of scaling altogether by
introducing deviations from linearity in the spin wave spectrum and
appearance of spectral weight in
modes not contained in the quantum nonlinear $\sigma$-model.
We argue that this is not the case here.
Indeed, from the fact that Eqs.(\ref{chi:lin:3},\ref{Schi:lin:3})
remain valid when $c(T)$ is already temperature dependent, we conclude
that the spin wave spectrum still remains linear even though
$c(T)$ deviates from its T=0 value. By the same token, we conclude that
there are no modes with significant spectral weight other than those
described by the $\sigma$-model.

Another universal quantity which has not been discussed in our paper
so far is the overall prefactor, $\chi_Q$,
in front of the scaling expression for
$\chi({\bf q},\omega)$ in Eq.(\ref{scaling:chi}).
This prefactor, which is renormalized by short wavelength quantum
fluctuations, can be expressed in terms of $N_0$ and $\rho_s$ for
temperatures where corrections of order $T/JS$ can be ignored. Where
$c$ is $T$-dependent, this quantity becomes $T$-dependent as well.
However, quantum corrections to the prefactor due to $1/S$
terms have not yet been calculated analytically.

Now the reason why the quantum critical scaling fails to
describe the temperature dependence of the correlation length
is clear. The correlation length
derived from Eq.(\ref{chiim}) is $\xi=c/m$. If one assumes
that the spin wave velocity is temperature independent above
$T=0.4J$, the quantum critical expression does not agree with the
numerical data. Given the discussions above, we argue that one should
consider not the temperature dependence of $\xi$, but that of
$m=c/\xi$, because $m$ is unaffected by the lattice corrections to $c$.
Then, the crossover from RC behavior to QC behavior
at temperatures around $J/2$ becomes evident.

\section{Conclusion} \label{conclusion}

Magnetic properties of the doped superconducting cuprates may be
closely related to the undoped parent compounds, such as
$\rm La_2CuO_4$ and $\rm Sr_2CuO_2Cl_2$,
which are well described by the square lattice Heisenberg model.
Motivated by this relationship, we have studied
scaling and the role of lattice
corrections in two dimensional square
lattice Heisenberg antiferromagnets.

In \cite{CHN}, Chakravarty, Halperin, and Nelson mapped 2D
collinear quantum antiferromagnets onto the quantum nonlinear $\sigma$-model,
and derived the corresponding phase diagram. When
the T=0 ground state of the corresponding $\sigma$-model has Neel order,
as is the case for the square lattice Heisenberg models,
the low temperature scaling regime is classical
(often called renormalized classical, or RC),
where the characteristic energy scale $\bar{\omega}\ll T$, while at
higher temperatures a crossover to the quantum critical (QC) regime
may be observed, where
$\bar{\omega}\sim T$. The S=1/2 square lattice
Heisenberg model, which describes $\rm La_2CuO_4$, was shown to exhibit
RC behavior at low temperatures \cite{CHN}.
As it has a substantial ordered moment,
it was earlier believed not to have a region of
quantum-critical behavior at intermediate
temperatures.

More recently, Chubukov and Sachdev \cite{CSY} argued that
the crossover from the RC to nonuniversal microscopic
behavior occurs through
the intermediate QC regime at $0.4J<T<0.6J$, with universal
behavior extending up to $T=0.6J$ and nonuniversal behavior above that
temperature. The importance of this at first sight  minor difference
concerning a fairly narrow temperature range, arises from the fact that
upon doping the QC region was argued to expand rather rapidly,
covering a much broader, and well accessible
 temperature range \cite{CSY}. This scenario  has important
implications for our views of magnetism in the
superconducting cuprate materials
\cite{CSY,Sokol:Pines}.

In our earlier paper \cite{Sokol:Glenister:Singh},
we showed that the scaling
functions for the equal time spin correlator, $S({\bf q})$,
and the static susceptibility, $\chi({\bf q})$, in the S=1/2 Heisenberg
model in the temperature range $T=0.6J-J$
agree with those calculated for the $\sigma$-model in the
QC regime. The same turned out to be true for the ratio
$S({\bf q})/T\chi({\bf q})$. Our comparisons did not involve any
adjustable parameters.

In the present paper, we undertook a detailed study of these issues using
the high temperature series expansions approach.
The simultaneous study of S=1/2 and S=1 models allows us to contrast the two
cases and highlight the
parameter-free
agreement of the intermediate temperature behavior
for the $S=1/2$ case with the
universal quantum critical
predictions of the non-linear $\sigma$-model.
For both $S=1/2$ and $S=1$ models, we find that the
q-dependences of $S({\bf q})$ and $\chi({\bf q})$ separately,
as well as of their ratio, agree very well up to $T\sim 0.6c_0/a$ with the
$\sigma$-model predictions calculated separately using $1/N$ expansions
(for $1/N$ calculations,
we utilized many of the results obtained in \cite{CSY}). This allows us
to calculate the temperature dependence of
the spin wave mass, defined as $m=c/\xi$, for which both RC and QC predictions
are known without adjustable parameters.

For S=1/2, we find that $m(T)$ is in good agreement with the RC
prediction for $T<0.45J$, while for
$0.65J<T<J$ the results are very close to the QC prediction and
far from the RC prediction, exhibiting
a crossover between these two regimes for $0.45J<T<0.65J$.
Furthermore, we find that for $T>0.5J$,
the lattice corrections cause a temperature dependence of
the spin-wave velocity $c$ and affect
the overall prefactor in front of the scaling expression for
$\chi({\bf q},\omega)$.
The new results strongly support our earlier conjecture
\cite{Sokol:Glenister:Singh}
that lattice corrections are such that
full q- and $\omega$-dependences
remain universal in the QC regime for temperatures up to $T\simeq
0.6c_0/a\sim J$:
\begin{equation}
\chi({\bf q},\omega) = \chi_Q \,
 \Phi^{\rm QC}\left(\frac{c|{\bf q}-{\bf Q}|}{T},\frac{\omega}{T}\right),
\label{characteristic:qc}
\end{equation}
where $\Phi^{\rm QC}(x,y)$ is the scaling function of the QC regime,
computable using,
for instance, the approach of \cite{CSY}.
For the S=1 model, where quantum fluctuations are substantially less, we find
that RC behavior directly crosses over to the high temperature interacting
local-spin behavior without an intervening QC regime.
Finally, the observed lattice corrections to $c$ for both S=1/2 and S=1
roughly agree in magnitude and temperature dependence with earlier
calculations of quantum $T/JS$ effects
by Kaganov and Chubukov \cite{Kaganov:Chubukov}.

In their recent publication, Greven and coworkers \cite{Greven} pointed out
the agreement of the measured and numerically calculated
correlation length $\xi(T)$
with RC prediction for all temperatures below $T=J$.
In our work, we present an alternative theory which is not only
consistent with the same
data for $\xi(T)$, but explains the q- and $\omega$-dependences
of spin correlators as well. We show that
detailed examination of the wavevector and temperature
dependence of $S(q)$ and $\chi(q)$ supports the picture of
renormalized classical behavior for temperatures up to $T=0.45J$, and
a crossover to a quantum critical behavior around $T=J/2$.

While the QC behavior for the spin wave mass occurs in our calculations in the
temperature range $T>0.6J$
currently inaccessible to experiments in La$_2$CuO$_4$ and
Sr$_2$CuO$_2$Cl$_2$,
many signatures of the QC behavior have been
experimentally observed at lower temperatures.
For instance, the nuclear relaxation rate, $1/T_1$,
in La$_2$CuO$_4$ saturates above
$T=600K\sim 0.4J$ \cite{Imai}
at a value close in magnitude to the universal QC
prediction \cite{CSY}. This is consistent with the present study,
as it has been shown \cite{CSY,Chubukov:Sachdev:Sokol}
that for those quantities which in the RC regime
depend on $\xi$ through its logarithm only, such as
the ratio $T_1T/T_{\rm 2G}$ and the bulk susceptibility, $\chi_0$,
the RC-QC crossover should be shifted substantially towards
lower temperatures, in agreement with
the measurements of $\chi_0$ \cite{Johnston}
and $T_1T/T_{2g}$ \cite{Imai} in La$_2$CuO$_4$.

It has been proposed earlier that the temperature
range of quantum critical behavior
rapidly expands with doping, and that the scaling behavior described by
the quantum nonlinear $\sigma$-model may be observed in doped antiferromagnets,
and specifically in the high temperature superconductors
\cite{CSY,Sokol:Pines,Sokol:Glenister:Singh}.
We hope that the study of the
renormalized classical to quantum critical scaling crossover and of the
origin of nonuniversal corrections to scaling in the insulator
presented here, will be helpful in understanding magnetic
behavior of the doped systems as well.

\acknowledgments

We are grateful to R.J. Birgeneau,
S. Chakravarty, A.V. Chubukov, M. Greven,
A.J. Millis, D. Pines, S. Sachdev, A.W. Sandvik, and
D.J. Scalapino
for many useful discussions. N.E. was supported by the
Deutsche Forschungsgemeinschaft.
R.L.G.\ and R.R.P.S.\ were supported by the NSF Grant DMR93-18537.
A.S.\ was supported by the NSF Grant
DMR89-20538 through the Frederick Seitz Materials Research Laboratory.
The computations have been performed using the IBM RISC/6000 workstation
received
through the Shared University Research Grant from the IBM Corporation.
N.E. would like to acknowledge the kind hospitality
of A.P.\ Young and Physics Department
of the University of California at Santa Cruz where part of this work
was done. A.S. is grateful to the Aspen Center for Physics for
hospitality during intial stages of this work.


\appendix

\section{Quantum Nonlinear $\sigma$-Model} \label{sigma:app}

The mapping \cite{CHN} of the low energy spectrum of the Heisenberg model
to the quantum nonlinear sigma model
has been extensively discussed in the literature and textbooks
(e.g., see \cite{Fradkin:book}). Here we
briefly discuss some aspects of this mapping, and of the $1/N$
expansion in this model,
as necessary for our purposes. We refer the reader
to Refs.\cite{CHN,CSY} for a far more detailed treatment.

The action of the quantum nonlinear $\sigma$-model can be written as:
\begin{eqnarray}
S = - \frac{\rho_{s0}}{2} \int d^2{\bf r} \int_0^{1/T} d\tau
\left[ \frac{1}{c_0^2} (\partial_\tau \vec n)^2 +
(\nabla \vec n)^2 \right],   \nonumber \\
\vec n^2 = 1,
\label{action}
\end{eqnarray}
where the constraint $ \vec n^2 = 1$ captures the scattering effects
and gradient terms lead to the linear spin wave spectrum. Here
$\vec n $ is the antiferromagnetic order parameter,
$\rho_{s0}$ is the bare spin stiffness and $c_0$ the bare
spin wave velocity. The assumed ultraviolet cutoff is set by the
lattice scale of the underlying Heisenberg model.

This model has been extensively studied in recent years
using renormalization group methods\cite{CHN},
Monte Carlo simulations\cite{Manousakis:Salvador},
and $1/N$ expansions \cite{CSY}.
The results obtained using different techniques agree with
each other quite well.
In our calculations, we rely mostly on the $1/N$ expansions
approach. In the next Subsection, we discuss the leading order of this
expansion, namely, the $N=\infty$ approximation, and then proceed with
the subleading $1/N$ corrections.

\subsection{N=$\infty$ Approximation} \label{N:infty}

The $N=\infty$ approximation is based on the replacement of the physical
$O(3)$ quantum nonlinear $\sigma$-model, where the three-component order
parameter $\vec n$ corresponds to three-component
physical spin, by the $O(\infty)$
$\sigma$-model where the number of components of $\vec n$ is infinite. A
consequence of such a replacement is the absence of damping.
This makes the $N=\infty$
approximation unsuitable for such problems as the nuclear spin-lattice
relaxation or the
neutron scattering, where correct description of damping is essential.
Nevertheless, for our particular purposes of extracting scaling
parameters from the data, already the $N=\infty$ approximation
is relatively accurate. It is further improved by taking into account the
first subleading corrections, which are discussed in the next Subsection.

The unit length constraint can be enforced
using Lagrange multiplier, $\lambda$:
$$
S =  - \frac{\rho_{s0}}{2} \int d^2{\bf r} \int_0^{1/T} d\tau
\hfill
$$
\begin{equation}
\times \left[ \frac{1}{c_0^2} (\partial_\tau \vec n)^2 +
(\nabla \vec n)^2
+ i\lambda \left( \vec n^2 -1 \right) \right].
\label{action:lambda}
\end{equation}
The action is now quadratic in $\vec n$ which can therefore
be integrated out, leaving
a functional integral for the field $\lambda$.
In case of $N=\infty$, the $\lambda$ field does not fluctuate
around the saddle point value $i\langle \lambda \rangle$, which in what
follows is denoted as
$m^2$ for the reason which becomes clear momentarily.
By performing Fourier transform of Eq.(\ref{action:lambda}),
one obtains the $N=\infty$ solution for the Matsubara frequencies
\cite{CSY}:
\begin{equation}
\chi_s(q,i\omega_n) = \frac{A}{m^2 + c^2 q^2 + \omega_n^2},
\ \ \ \xi=\frac{c}{m},
\label{chiim}
\end{equation}
where $c$ is the spin wave velocity and it is now evident that
$m=c/\xi$ has the meaning of the spin wave mass,
i.e.\ the gap in the spin wave spectrum.

The Matsubara
correlator $\chi_s (q, i \omega_n )$, where $\omega_n = 2 \pi n T$,
is related to the physical real frequency response function by
analytical continuation: $\omega \to i\omega_n$. It is easy to see
that Eq.(\ref{chiim}) corresponds to a spectrum of undamped
magnons with a gap:
\begin{equation}
\chi_s(q,\omega) = \frac{A}{m^2 + c^2 q^2 - \omega^2} =
\frac{A}{2\epsilon_q} \left[ \frac{1}{\epsilon_q-\omega} +
\frac{1}{\epsilon_q + \omega} \right],
\end{equation}
where the spin wave dispersion is:
\begin{equation}
\epsilon_q = \left(m^2 + c^2 q^2 \right)^{1/2}.
\end{equation}
The imaginary part of $\chi(q,\omega)$, which describes dissipation,
is:
\begin{equation}
\chi_s''(q,\omega) = \frac{A}{2\epsilon_q} \left[
\delta(\omega - \epsilon_q) - \delta(\omega + \epsilon_q) \right]
\ \ \ \ {\rm at} \ \ \ \ N=\infty.
\end{equation}
Of course, for the physical $O(3)$ QNL$\sigma$ model the damping
is finite and the peaks in $\chi''$ around
$\omega = \pm \epsilon_q$ acquire finite width. This effect is captured
only in the terms beyond the $N=\infty$ approximation.

Our considerations for the S=1/2 and S=1 Heisenberg models are based
on the high temperature series data for the
q-dependent static susceptibility of the staggered order parameter ($\vec
n$) field , $\chi_s(q)$,
and equal time structure factor of the same field, $S(q)$.
Using Eq.(\ref{chiim}), one obtains:
\begin{equation}
 \chi(q) = \chi_s(q,\omega_n=0) = \frac{A}{\epsilon_q^2},
\label{static.chi.infty}
\end{equation}
\begin{equation}
S(q) = T \sum_{n=-\infty}^{+\infty} \chi_s(q,i\omega_n) =
\frac{A}{\epsilon_q}
\tanh^{-1} \left[ \frac{\epsilon_q}{2T} \right],
\label{S.infty}
\end{equation}
where
\begin{equation}
A = \chi_s(0)\times m^2,
\ \ \ \
\epsilon_q = \left(m^2 + c^2 q^2\right)^{1/2}.
\label{A.epsilon}
\end{equation}
It follows directly from Eq.(\ref{static.chi.infty}) that:
\begin{equation}
\frac{\chi_s(0)}{\chi_s(q)} =
1 + \left(\frac{c}{m}\right)^2 q^2 \ \ \ \mbox{for $N=\infty$}.
\label{chi:lin:infty}
\end{equation}
Further, the ratio of Eq.(\ref{static.chi.infty})
and Eq.(\ref{S.infty}) depends only on $\epsilon_q/T$:
\begin{equation}
	\frac{S(q)}{T\chi_s(q)} = \frac{\epsilon_q}{2T}
	 \tanh^{-1} \left[  \frac{\epsilon_q}{2T} \right].
	\label{ratio:infty}
\end{equation}
We now introduce a function $\gamma[x]$ defined as:
\begin{equation}
\gamma\left[\frac{x/2}{\tanh(x/2)}\right] \equiv x,
\ \ \ \ {\rm for} \ \ \ \ x>1,
\label{gamma:def:2}
\end{equation}
which has the following asymptotic
behaviors: $\gamma[x]\simeq 2 \sqrt{3(x-1)}$
for $0<x-1\ll 1$ and $\gamma[x]\simeq 2x$ for $x\gg 1$.
Then
\begin{equation}
\gamma^2 \left[ \frac{S(q)}{T\chi_s(q)} \right] =
\left( \frac{m}{T} \right)^2 + \left( \frac{c}{T} \right)^2 q^2
\ \ \ \mbox{for $N=\infty$}.
\label{Schi:lin:infty}
\end{equation}

\subsection{$1/N$ Expansion} \label{1/N}

In this Subsection, we discuss corrections to the $N=\infty$ equations
(\ref{chi:lin:infty},\ref{Schi:lin:infty}) for the physical case of $N=3$.
We utilize the results of the $1/N$ expansion calculations
by Chubukov, Sachdev, and Ye \cite{CSY}
and find that Eqs.(\ref{Schi:lin:infty},\ref{chi:lin:infty})
remain quite accurate even for $N=3$.

We turn first to Eq.(\ref{Schi:lin:infty}). While
the full q-dependence of $1/N$
corrections have not been calculated so far, the results for small
wavevectors as presented in \cite{CSY} turn out to be quite sufficient for
our purposes.
In the asymptotic high temperature (quantum critical) regime,
$m/T$ is temperature independent, while $\chi_s(0)/\chi_s(q)$ and
$S(q)/T\chi_s(q)$ depend only on the product $ q\xi \sim cq/T $.
With straightforward application of
the results of \cite{CSY}, we obtain for small $q$:
\begin{equation}
\frac{\chi_s(0)}{\chi_s(q)} = 1 + \left( v_1 \frac{c}{m} \right)^2 q^2
\label{chi:lin:3:app}
\end{equation}
\begin{equation}
\gamma^2 \left[ \frac{S(q)}{T\chi_s(q)} \right] =
\left( u_2 \frac{m}{T} \right)^2 + \left( v_2 \frac{c}{T} \right)^2 q^2,
\label{Schi:lin:3:app}
\end{equation}
where
\begin{eqnarray}
v_1 &= 1 - 0.0042/N &\to 0.9986, \nonumber \\
u_2 &= 1 + 0.1473/N &\to 1.0491, \nonumber \\
v_2 &= 1 + 0.0382/N &\to 1.0127
\label{uv:app}
\end{eqnarray}
($1/N$ terms are evaluated for $N=3$).

The $1/N$ corrections to the coefficients $v_1$, $v_2$,
which enter in parenthesis in front of $q^2$ in
Eqs.(\ref{chi:lin:3:app},\ref{Schi:lin:3:app}), are essentially negligible.
Higher terms in the expansion in powers of $q^2$ have not been
calculated so far. However, once the leading terms have very small
corrections, corrections to higher order terms in the expansion in $q^2$
should be of the same order of magnitude. The reason for this conclusion
is as follows.
Eqs.(\ref{chi:lin:3:app},\ref{Schi:lin:3:app}) can be considered as
relationshps
between the position of the closest to the origin singuliarity for
imaginary wavevectors, $q_{\rm sing}=1/\xi$, and the expansion in powers
of $q^2$ near $q=0$. Were the higher order terms large, the position of
the singuliarity expressed in terms of the coefficient in front of $q^2$,
namely, $v_1$ or $v_2$, would change, which in its turn would lead to
large corrections to $v_1, v_2$. Given the fact that such corrections are
in fact small, and ignoring the possibility of accidental cancellations
for both Eqs.(\ref{chi:lin:3:app},\ref{Schi:lin:3:app}), we arrive to the
conclusion that higher order $q^{2n}$ terms are small and therefore
Eqs.(\ref{chi:lin:3:app},\ref{Schi:lin:3:app}) are applicable for
$qa\ll 1$, not only for the much narrower wavector range $q\xi\ll 1$.

It is quite remarkable that the $1/N$ correction to the coefficient $u_2$
is small even when evaluated at $N=3$. Indeed, the ratio
$S({\bf Q})/T\chi({\bf Q})$
reflects the distribution of the spectral weight over
frequencies. At the commensurate wavevector, the character of this
distribution changes from a delta-function peak at finite $\omega=m$ to a
smooth peak centered at $\omega=0$ as one goes from $N=\infty$ to $N=3$,
which in general could have led to large $1/N$ corrections to $u_2$.
The fact that the rigorously calculated correction (\ref{uv:app}) is in fact
small shows that already the $N=\infty$ approximation accurately
captures the relationship
between the spin wave mass and the relative correlation strength at
different Matsubara frequencies.

Further, since the above corrections are essentially negligible
in the QC regime where they have been calculated, it iseems reasonable to
conjecture that they may not
increase drastically as $m/T$ decreases, in which case they should apply
to the RC-QC crossover regime as well. The rigorous calculations to
check this conjecture are prohibitively complicated and therefore have
not been carried out. Instead, in the next Section we verify
Eqs.(\ref{chi:lin:3:app},\ref{Schi:lin:3:app})
are in agreement with the numerical
data for Heisenberg models
for all temperatures studied,
and thereby establish that the above conjecture is indeed correct, which
greatly simplifies futher considerations.

To summarize, in this Appendix we derived surprisingly simple
Eqs.(\ref{chi:lin:3:app},\ref{Schi:lin:3:app},\ref{uv:app}).
Negligible (for $v_1$, $v_2$) or small ($\sim 5\%$ for $u_2$)
$1/N$ corrections to the $N=\infty$ result indicate a well behaved
large $N$ expansion, and we expect the above equations
to be accurate for the physical $O(3)$ model.
The correspondence between the Heisenberg and
sigma-models is such that
$\chi_s(q,\omega)$ for the $\sigma$-model
is  $\chi(|{\bf q}-{\bf Q}|,\omega)$ for the Heisenberg model,
where ${\bf Q}=(\pi,\pi)$. Taking this into account,
we rewrite Eqs.(\ref{chi:lin:3:app},\ref{Schi:lin:3:app})
in the form appropriate
for the Heisenberg models, Eqs.(\ref{chi:lin:3},\ref{Schi:lin:3}),
and use them in Section \ref{universal}
for comparisons with the numerical data.

\section{Correlation Length} \label{xi:app}

\begin{figure}
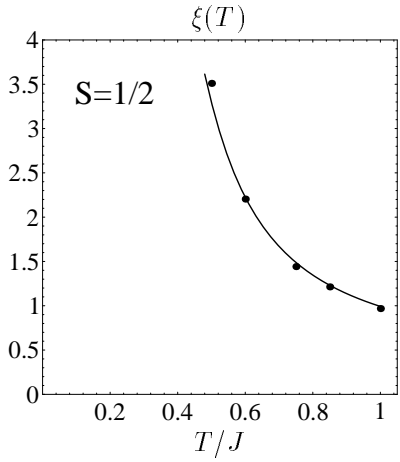

\caption{
The correlation length determined using our new method of analytical
continuation to imaginary wavevectors (solid line) plotted along with
earlier Monte Carlo results or Makivic and Ding
\protect\cite{Ding:Makivic} (circles).}
\label{xi.vs.monte.carlo:fig}
\end{figure}

In this Appendix, we verify our resluts for $m$ and $c$ by comparing the
ratio $c/m$ with the independently calculated correlation length.
We calculate the correlation length, defined as the inverse
rate of exponential decay of spin correlator at large distances,
directly using our new ``imaginary wavevector'' method
\cite{Glenister:Elstner:Singh:Sokol}. The idea of this calculation
is based on an observation that when the pair
correlation function decays exponentially up to power-law corrections,
which is the case for two-dimensional Heisenberg models,
the Fourier transform of spin correlator around the ordering wavevector,
$\chi({\bf Q}+{\bf n}_q q)$, where ${\bf n}_q$ is a unit vector,
should have a singuliarity on the imaginary
$q$ axis at $q_{\rm sing} = i/\xi$ when analytically continued to
complex $q$. The position of this
singuliarity can be determined through high temperature series expansions
using the methods described in
\cite{Glenister:Elstner:Singh:Sokol}.
In Fig.\ref{xi.vs.monte.carlo:fig}, our results for $\xi$
for the S=1/2 Heisenberg model are plotted along with the earlier quantum
Monte Carlo calculations by Makivic and Ding \cite{Ding:Makivic};
the agreement is excellent at all temperatures.

\begin{figure}
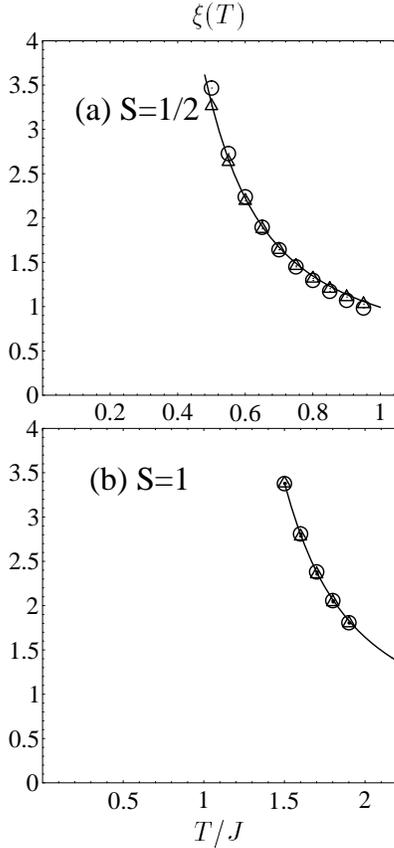

\caption{
The correlation length for the S=1 Heisenberg model
calculated directly using the analytical
continuation to imaginary wavevectors (solid line), and independently
using the mapping to the quantum nonlinear $\sigma$-model, from
Eq.(\protect\ref{chi:lin:3}) (triangles) and
Eq.(\protect\ref{Schi:lin:3}) (circles).
All three sets of data agree with each other for both S=1/2 and S=1.}
\label{xi:comp:fig}
\end{figure}

Now we can verify the equality $\xi=c/m$, which holds by the definition
of $m$. In Fig.\ref{xi:comp:fig}, we
plot our directly calculated $\xi$ by a solid line,
the ratio $c/m$ determined solely from Eq.(\ref{chi:lin:3})
as hollow circles, and the ratio of $c$ and $m$ with both
determined seperately from Eq.(\ref{Schi:lin:3}) as solid circles.
The agreement between all three sets of data is excellent for both S=1
and S=1/2. This agreement not only indicates that our calculations of
$m$ and $c$ are accurate, but also serves as another
verification
that the $\sigma$-model corectly describes wavevector and frequency
dependences of the dynamical susceptibility of the Heisenberg models
in the studied temperature range.


\begin{references}

\bibitem{reviews}
For comprehensive reviews, see
S. Chakravarty, in Proceedings of
{\em High Temperature Superconductivity}, edited by K.S. Bedell
{\em et al.} (Addison-Wesley, CA, 1990), and
E. Manousakis, Rev. Mod. Phys. {\bf 63}, 1 (1991).

\bibitem{Sandvik:Scalapino:PRL}
A.W. Sandvik and D.J. Scalapino, Phys. Rev. Lett., {\bf 72}, 277 (1994),
and unpublished.
The two plane model was discussed earlier
by K. Hida, J. Phys. Soc. Japan, {\bf 61}, 1013 (1992);
A.J. Millis and H. Monien, Phys. Rev. Lett., {\bf 70}, 2810 (1993).

\bibitem{CHN}
S. Chakravarty, B.I. Halperin, and D.R. Nelson,
Phys. Rev. B {\bf 39}, 2344 (1989); see also
S. Chakravarty and R. Orbach, Phys. Rev. Lett. {\bf 64}, 224 (1990).

\bibitem{Hasenfratz:Niedermayer}
P. Hasenfratz and F. Niedermayer, Phys. Lett. B {\bf 268},
231 (1991); Z. Phys. B {\bf 92}, 91 (1993).

\bibitem{CSY}
A.V. Chubukov and S. Sachdev, Phys. Rev. Lett. {\bf 71},
169 (1993);
A.V.\ Chubukov, S.\ Sachdev, and J.\ Ye, Phys. Rev. B{\bf 49},
11919 (1994).

\bibitem{Johnston} D.C. Johnston, Phys. Rev. Lett. {\bf 62}, 957 (1989);
D.C. Johnston, S.K. Sinha, A.J. Jacobson and
J.M. Newsam, Physica (Amsterdam),  {\bf 153-155 C}, 572 (1988)

\bibitem{Imai}
T. Imai, C.P. Slichter, K. Yoshimura and K. Kosuge,
Phys. Rev. Lett. {\bf 70}, 1002 (1993);
T. Imai, C.P. Slichter, K. Yoshimura, M.Katoh and K. Kosuge,
Phys. Rev. Lett. {\bf 71}, 1254 (1993).

\bibitem{Greven}
M. Greven {\em et al.},
Phys. Rev. Lett. {\bf 72}, 1096 (1994).

\bibitem{Keimer}
B. Keimer {\em et al.}, Phys. Rev. B {\bf 46}, 14034 (1992).

\bibitem{Ding:Makivic} H.-Q. Ding and M.S. Makivic, Phys. Rev. Lett,
{\bf 64}, 1449 (1990); M.S. Makivic and H.-Q. Ding,
Phys. Rev. B {\bf 43}, 3662 (1990).

\bibitem{Singh:Glenister}
R.R.P. Singh and R.L. Glenister,
Phys. Rev. B {\bf 46}, 11871 (1992).

\bibitem{Sokol:Glenister:Singh}
A. Sokol, R.L. Glenister and R.R.P. Singh, Phys. Rev. Lett., {\bf 72},
1549 (1994).

\bibitem{Manousakis:Salvador}
E. Manousakis and R. Salvador, Phys. Rev. Lett., {\bf 62}, 1310 (1989);
Phys. Rev. B {\bf 40}, 2205 (1989). Remark: in our opinion, the
data for the slab thickness $N_\tau=2$ may not be appropriate for extracting
properties of the thermodynamic limit ($N_\tau=\infty$),
where $m/T$ is defined. We
therefore repeated the analysis of Manousakis and Salvador omitting the
data for $N_\tau=2$, arriving to the errorbars for $m/T$, shown in
Eq.(\ref{m:all.methods:QC}),
which are larger than the errorbars quoted in the original
paper. With this larger erorbars, the result agrees with
calculations in the framework of other approaches.

\bibitem{Singh:rhos}
R.R.P. Singh, Phys. Rev. B {\bf 39}, 9760 (1989);
R.R.P. Singh and D. Huse, Phys. Rev. B {\bf 40}, 7247 (1989).

\bibitem{Wiese:Ying}
U.J. Wiese and H.P. Ying, Z. Phys. B{\bf 93}, 147 (1994).

\bibitem{1/S}
T. Oguchi, Phys. Rev, {\bf{117}}, 117 (1960); S. Chakravarty
and C. Castilia, Phys. Rev. B {\bf 43}, 13687 (1991);
J. Igarashi, Phys. Rev. B {\bf 46}, 10763 (1992).

\bibitem{Chubukov:Sandvik}
A.V. Chubukov and A.W. Sandvik, unpublished.

\bibitem{Singh:c}
R.R.P. Singh, Phys. Rev. B{\bf 47}, 12337 (1993).

\bibitem{Kaganov:Chubukov}
M.I. Kaganov and A.V. Chubukov, in ``Spin waves and Magnetic Excitations'',
part I, Eds. A.S. Borovik-Romanov and S.K. Sinha, North-Holland,
Amsterdam (1988).

\bibitem{Sokol:Pines}
A. Sokol and D. Pines, Phys. Rev. Lett. {\bf 71}, 2813 (1993).

\bibitem{Chubukov:Sachdev:Sokol}
A.V. Chubukov, S. Sachdev, and A. Sokol,
Phys. Rev. B {\bf 49}, 9052 (1994).

\bibitem{Fradkin:book}
E.H. Fradkin, Field Theories of Condensed Matter Systems, Addison-Wesley,
Redwood City, CA (1991).

\bibitem{Glenister:Elstner:Singh:Sokol}
R.L. Glenister, N. Elstner, R.R.P. Singh, and A. Sokol,
to be published.

\end{references}
\end{document}